\documentclass[a4paper,fleqn]{cas-sc}
\usepackage[numbers]{natbib}
\usepackage{float}
\usepackage[justification=centering]{caption}
\usepackage{graphicx}
\usepackage[font={footnotesize}]{caption}

\begin{document}
\shorttitle{Estimation of Acetabular Version from Anteroposterior Pelvic Radiograph Employing Deep Learning}
\shortauthors{Ata Jodeiri et~al.}

\title [mode = title]{Estimation of Acetabular Version from Anteroposterior Pelvic Radiograph Employing Deep Learning}

\author[1]{Ata Jodeiri}
\ead{ata.jodeiri@tabrizu.ac.ir}
\cormark[1]
\cortext[cor1]{Corresponding author: ata.jodeiri@tabrizu.ac.ir, Image Processing and Machine Vision Lab, Faculty of Electrical and Computer Engineering, University of Tabriz, 29 Bahman Blvd., Tabriz, Iran.\\
\hspace*{5mm}
*Corresponding author2: Seyyedhossein Shafiei, Orthopedic Surgery Research Centre, Sina University Hospital, Tehran University of Medical sciences, Tehran, Iran.}

\author[1]{Hadi Seyedarabi}
\ead{seyedarabi@tabrizu.ac.ir}
\address{$^{a}$Faculty of Electrical and Computer Engineering, University of Tabriz, Tabriz, Iran}

\author[2]{Fatemeh Shahbazi}
\ead{fshahbazy@gmail.com}
\address{$^{b}$School of Electrical \& Computer Engineering, University of Tehran, Tehran, Iran}

\author[3]{Seyed Mohammad Mahdi Hashemi}
\ead{smm-Hashemi@student.tums.ac.ir}

\author[3]{Seyyedhossein Shafiei}
\ead{dr_hshafiei@yahoo.com}
\address{$^{c}$Orthopedic Surgery Research Centre, Sina University Hospital, Tehran University of Medical sciences, Tehran, Iran}
\cormark[1]
\begin{abstract}
Background and Objective: The Acetabular version, an essential factor in total hip arthroplasty, is measured by CT scan as the gold standard. The dose of radiation and expensiveness of CT make anterior-posterior pelvic radiograph an appropriate alternative procedure. In this study, we applied a deep learning approach on anteroposterior pelvic X-rays to measure anatomical version, eliminating the necessity of using Computed tomography scan.\\
Methods: The right and left acetabular version angles of the hips of 300 patients are computed using their CT images. The proposed deep learning model, Attention on Pretrained-VGG16 for Bone Age, is applied to the AP images of the included population. The age and gender of these people are added as two other inputs to the last fully connected layer of attention mechanism. As the output, the angles of both hips are predicted.\\
Results: The angles of hips computed on CT  increase as people get older with the mean values of 16.54 and 16.11 (right and left angles) for men and 20.61 and 19.55 for women in our dataset. The predicted errors in the estimation of right and left angles using the proposed method of deep learning are in the accurate region of error (<=3 degrees) which shows the ability of the proposed method in measuring anatomical version based on AP images.\\
Conclusion: The suggested algorithm, applying pre-trained vgg16 on the AP images of the pelvis of patients followed by an attention model considering age and gender of patients, can assess version accurately using only AP radiographs while obviating the need for CT scan. The applied technique of estimation of anatomical acetabular version based on AP pelvic images using DL approaches, to the best of authors' knowledge, has not been published yet.
\end{abstract}


\maketitle

\section{Introduction}\label{int}
\setlength{\baselineskip}{1.5\baselineskip}
The Acetabular version, defined as operative anteversion, sharp angle, and anatomical anteversion \cite{ref:1}-\cite{ref:2}, refers to the orientation related to the opening of the acetabulum \cite{ref:3}. These definitions are based on the measurements performed on the sagittal, coronal, and horizontal planes \cite{ref:1}-\cite{ref:2}. Natural anatomical anteversion is 20.7 $\pm$ 3.8 \cite{ref:4}, while it is recommended for component positioning between $0^{\circ}$ and $30^{\circ}$ in various studies \cite{ref:5} and  suggested by Lewinnek et al. to be $15^{\circ}$ \cite{ref:5}. Anteversion is defined as an important changing morphological feature in the biomechanics \cite{ref:6} and pathologies of the hip joint \cite{ref:7}. The motion of the hip in both internal and external is proved to be affected by femoral and acetabular versions \cite{ref:6}. Version abnormalities are also shown to be associated with hip pain \cite{ref:6}, acetabular fractures, osteoarthritis \cite{ref:7}, dysplasia, and femoroacetabular impingement \cite{ref:8}. Moreover, in hip surgeries such as total hip arthroplasty (THA), which is an attempt to soothe pain, improve functional activity, and enhance quality of life \cite{ref:9}, determination of acetabulum version is a crucial factor in both predicting and following up the outcome of the operation. Position of the cup in THA is a significant issue in which inappropriate positioning entails impingement, dislocation, severe wear, and reduction in the range of motion \cite{ref:9}-\cite{ref:11} which may force patients for revision surgery \cite{ref:10}. Assessment of the orientation of acetabulum has been done by 3-D imaging systems (CT-scan and MRI) and anterior-posterior (AP) pelvic X-rays \cite{ref:11}.

Klasan et al. in 2019 studied the whole-body CT scans of 404 patients (aged 16 to 40 years) and reported a higher average acetabular version in women (19.31 $\pm$ 5.04 compared to 16.46 $\pm$ 4.42 in men) while the version increases with age. The anatomical measurement was done on cranial, central and caudal levels \cite{ref:7}. In 2018, in a study by Lerch et al. \cite{ref:6}, femoral and acetabular versions of 462 symptomatic patients with hip pain were examined between desired subgroups and compared with a control group. The results showed significant differences in acetabular version in some of the subgroups compared to the control group. The researchers in this study observed that the femoral version's abnormalities in the subjects suffering from hip pain, who are qualified for hip preservation surgery, were more prevalent than in the control group.  As a result, it is recommended that the femoral version and acetabular version should be assessed in youth with painful hips using CT or MRI imaging techniques. A comparison was made based on the assessment of anteroposterior radiograph of the selected cohort using computer software approach and CT of the whole body of the control subjects. Two different acetabular version computing radiographic methods, ischiolateral and Woo \& Morrey, were conducted by Raj et al. \cite{ref:9} on the AP images of patients with Total Hip Replacement (THR) surgery. Comparing the results with the gold standard CT using regression line, the values of Woo \& Morrey were reported close to CT values except in hip stiffness in the contralateral hip. AP radiograph of hip and pelvis of 153 dysplasia patients were examined in order to assess the prevalence of acetabular retroversion \cite{ref:3}. Snijders et al. \cite{ref:11} measured radiographic anteversion based on ten different mathematical approaches using patients' AP pelvic radiography. Finally, they compared the results of these methods with the gold standard on CT images. The linear correlation of the measurements was in the [0.556$\:$ 0.747] interval in which CT scan was recommended as the reliable method for anteversion measurement \cite{ref:11}. Lots of researchers evaluated cup versions to assess the influence of cup malposition in THA surgery using CT or lateral-pelvis radiographs \cite{ref:10}. In some studies, lateral radiographs were proposed for anteversion calculations though the others did not acknowledge them in distinguishing anteversion from retroversion. Finally, the version was recommended to be measured accurately using standard CT by many scholars.

Though CT is introduced as the gold standard method of acetabular version measurement \cite{ref:9}, AP radiograph images are the first assessment in patients with pain in the hip \cite{ref:8} due to their availability, less radiation, and cost-effectiveness \cite{ref:9}-\cite{ref:10}. Typically, 20$\%$ to 50$\%$ of high-tech imaging methods, including CT, are declared unnecessary \cite{ref:12}, where considering the dose of radiation in CT, anteroposterior pelvis radiographs can be considered as an appropriate alternative procedure. As a result, a low-risk and low-cost method for accurate assessment of version is regarded as a significant need. In this study, we used AP radiographs for assessment of the orientation of acetabulum. 

Here we propose an algorithm based on artificial intelligence (AI) in the anatomical acetabular version measurement using AP images. New technologies in orthopedics have received much attention in such a way that we can see considerable developments in preoperative training, preoperative planning, and intraoperative navigation to assist orthopaedists \cite{ref:13}. Nowadays, AI has attracted more attention using deep learning (DL) as a class of machine learning (ML) applications in the diagnosis of medical images \cite{ref:14}-\cite{ref:15}. Radiological image interpretation and diagnoses applying ML method have been considered recently \cite{ref:16}. ML has also been applied in different fields, including: segmentation of pulmonary embolism using angiography, colon polyp detection with CT scans, diagnosing and detecting breast cancer in mammography \cite{ref:16}.  Diagnosis of fractures and estimation of pediatric bone age are among the cases in that DL has been proved to be capable in orthopedics \cite{ref:14}. Jodeiri et al., 2020, proposed a deep learning approach to estimate pelvis sagittal inclination. His CNN-based method on the AP images of 475 patients with THR surgery, eliminated the necessity for taking CT scans \cite{ref:17}.

Convolutional neural networks (CNN) are a well-known architecture of DL, which is severely used in radiology \cite{ref:15}. Many studies have used CNN-based procedures in musculoskeletal radiology listed as lesion detection (fractures, cartilage abnormalities, ACL, and meniscal tears), classification (osteoarthritis quantification, grading of spinal stenosis, and bone age assessment), and segmentation \cite{ref:15}. Among the different methods of DL, VGG-16 was used in various applications as in detecting intertrochanteric hip fractures \cite{ref:18}, vertebral body localization (L3) \cite{ref:19}, determining sex by hand radiographs \cite{ref:20}, and segmentation of knee MRI \cite{ref:21}. VGG16 network, a CNN used in 2014 competition of ImageNet Large Scale Visual Recognition Competition (ILSVRC), is trained on ImageNet \cite{ref:16} (a 1000 class labels of 1.28-M training images).

In this study, our objectives, besides introducing a deep learning approach for estimation of anatomical acetabular version using AP X-rays, are summarized as follows:  1) to evaluate the correlation of acetabulum angle determined by the proposed method and by CT-scan as the gold standard, 2) to determine the mean value of acetabular version in the study population based on their age, 3) to determine the mean value of acetabular version in the study population based on their gender, and 4) to determine the mean value of acetabular version in the study population. Applying DL method in AP radiographs to estimate the anatomical version of the acetabulum, to the best of authors’ knowledge, is the first study of this kind. 

The methodology section, section \ref{met}, investigates AP pelvis radiographs and CT scans of 300 patients, the algorithm of DL method, statistical analysis, and cross-validation strategy. The achieved results of the above-mentioned purposes using DL approach are reported in section \ref{res}. Discussion and conclusion are the last parts in our study represented in sections \ref{dis} and \ref{con}, respectively.

\section{Materials and Method}\label{met}
\subsection{Dataset}\label{data}
\begin{figure}[!ht]
	\centering
	\captionsetup{justification=centering}
	\includegraphics[scale=1]{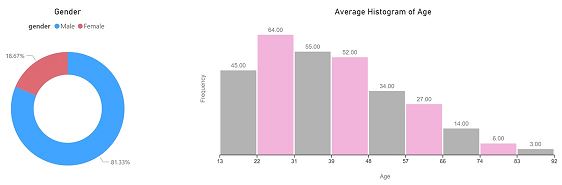}
	\caption{The distribution of the studied population according to their sex \& the histogram of the distribution for the population's age.}
    \label{pic:GenAge}
\end{figure}

In order to collect our dataset, we searched the patient archiving and communications system (PACS) of Sina University Hospital (Tehran University of Medical Sciences, Tehran, Iran) for multiple trauma-listed patients retrospectively. The Patients whose anterior-posterior pelvic (supine) radiographs along with abdominopelvic CT scans were available in the PACS system had been selected. The selected cohort contains 300 subjects (244 men and 56 women ranged 13 to 92 years old) who aged $\geq$ 13 years and their both AP radiographs and abdominal and pelvic CT scans were recorded simultaneously in Sina Hospital’s emergency room during the desired time of September 2018-September 2020. We excluded subjects who: i) their CT images showed fractures in their acetabulum, femoral head and neck, and pelvic ring, ii) had previous surgery or fracture of the pelvic bone or proximal femur, and iii) their AP pelvic images were of poor quality or non-standard (including tilt or rotation). The distribution of the studied population according to their sex together with the histogram of the distribution for the population's age are presented in Fig.\ref{pic:GenAge}. In tables \ref{tab:gender} and \ref{tab:age}, the information of angle of both hips for both sexuality populations and three age groups studied in our research are reported, respectively.

\begin{table}[!t]
	\caption{The information of angle of both hips for both sexuality populations.}
	\label{tab:gender}
	\begin{tabular*}{\tblwidth}{@{} |L|cccc|@{} }
		\hline
		Gender  & n    & age               & Right angle     & Left angle\\
		\hline
		Male    & 244  & 37.72$\pm$16.01   & 16.54$\pm$5.28  & 16.11$\pm$5.43\\
		Female     & 56   & 46.95$\pm$16.39   & 20.61$\pm$5.70  & 19.55$\pm$5.20\\
		\hline
		All & 300  & 39.44$\pm$16.47   & 17.30$\pm$5.58  & 16.75$\pm$5.55\\
		\hline
	\end{tabular*}
\end{table}

\begin{table}[!t]
	\caption{The information of angle of both hips for three age groups.}
	\label{tab:age}
	\centering
	\begin{tabular*}{\tblwidth}{@{} |L|ccc|@{} }
		\hline
		Age (years) & n    & Right angle      & Left angle\\
		\hline
		$\leq$45    & 191  & 16.10$\pm$5.33   & 15.50$\pm$5.30\\
		45-65       & 75   & 19.09$\pm$5.39   & 18.36$\pm$5.30\\
		$\geq$65    & 34   & 20.06$\pm$5.52   & 20.23$\pm$5.15\\
		\hline
	\end{tabular*}
\end{table}

The technique used for anterior-posterior pelvic radiographs is a standard view \cite{ref:22}, acquired in the supine position, at 120 cm as the distance of tube-to-image and a photon beam centered midway between the pubic symphysis and the top of the iliac crests without any rotation in the patient’s pelvis. A Philips Ingenuity Flex 16 slice CT-scan system was used to acquire spiral CT images of cases in the supine position, and 1.5 mm slices in axial plane were used. A medical student under direct supervision of a hip surgeon (SHS) estimated hip angles of both sides using CT scans of patients and recorded them as the gold standard.

\subsection{Model}\label{NNs}
The Attention on Pretrained-VGG16 for Bone Age model \cite{ref:23} proposed by Mader is the DL approach used in our study. He added an attention mechanism with convolutional layers to the pre-trained VGG16 \cite{ref:24}. Convolution layers, assuming as feature extractors, are the main blocks in CNNs \cite{ref:19}. When we stack several convolution layers, each of them extracting new features from previous ones, the hierarchical features from our input will be built \cite{ref:19},\cite{ref:25}. This particular architecture is reported as a success in lots of tasks in computer vision \cite{ref:25}. 

Patient privacy regulations and highly trained human experts for image annotation make medical images rare and inaccessible in order to train neural networks \cite{ref:25}. To overcome this challenge, the transfer learning has become the center of attention \cite{ref:19},\cite{ref:25}. So the model, VGG16 here, is trained using a vast labeled dataset, ImageNet in this model, attention mechanism is added and then fine-tuned using medical images to be adapted for our target of acetabular version estimation. The good performance of this solution, training on a vast labeled dataset and fine-tuning using medical images, is reported in \cite{ref:26}-\cite{ref:27}. Moreover, data shortage as well as model generalization are also addressed using different techniques of data augmentation similarly height and width shifts and zooming, all with the range of 0.05 and the filling mode is set to nearest \cite{ref:17}, all are performed during the training stage. For our purpose, rotation cannot be used as an augmentation technique since right and left angles of hips may be considered inaccurately. We also mirror our AP images as well as the angles. This approach is performed for the training subset in which during the training step, one of the original AP or the mirrored ones for each subject is selected randomly. 

\begin{figure}[!ht]
	\centering
	    \includegraphics[scale=0.7]{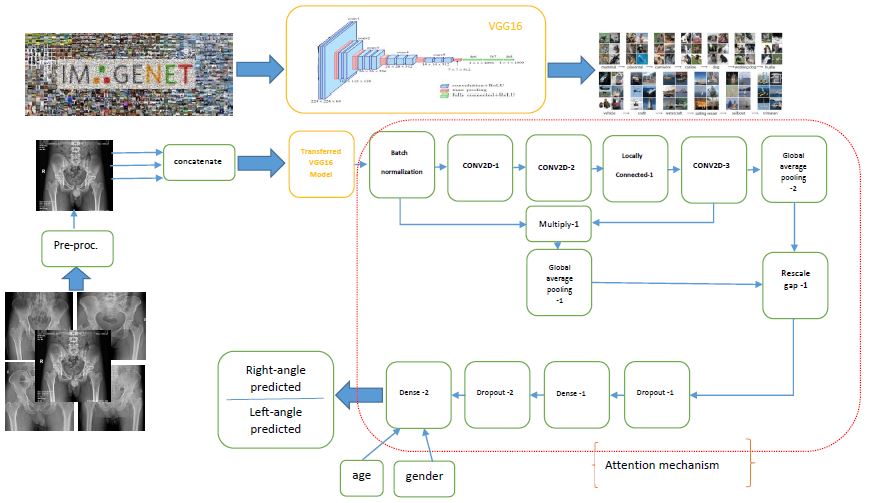}
	\caption{Block diagram of the proposed model.\label{pic:dig}}
\end{figure}

As our AP radiographs are in different sizes we rescale them to 1024$\ast$1024 in the preprocessing stage and then perform normalization. These grayscale radiographs are sent to the model as the input. In the next stage we concatenate three of each input image to make a $1024\ast1024\ast3$ image for the input to CNN of pre-trained VGG16. As shown in the diagram of the study (Fig.\ref{pic:dig}), batch normalization is done after performance of VGG16. The attention mechanism is added since the whole model focuses on the most relevant regions of radiographs. Before pooling, the gap’s pixels become on/off, followed by the lambda layer to rescale the pixels. The outputs, right and left angles, are then max normalized. At this point, the gender and age of patients are as two other inputs sent to two different neurons, with gender as 1 for male and 0 for female and ages of subjects are normalized between 0-1. At the last stage, the right and left angles are predicted. It is worth mentioning that for the first time, here the predicted output is anatomical version, and not computed from radiograph images.

Our model is trained over 1000 epochs. Every 30 epochs, if the loss function doesn’t decrease meaningfully, the learning rate (lr) would decrease from the base value ($10^{-3}$) by a factor of 0.8. At the end of the training stage, we observed that lr reached to $10^{-4}$.  Nesterov-accelerated Adaptive Moment, NADAM is selected as the optimizer with mean square error (MSE) as loss function. Employing an optimization algorithm (NADAM) explained as a kind of improved ADAM (Adaptive Movement Estimation) which adds NAG (Nesterov’s Accelerated Gradient), sometimes called Nesterov momentum, is considered to be an improved type of momentum presented by Dozat \cite{ref:28}.

\subsection{Cross-Validation}
Here, in order to assess the performance of the proposed algorithm, the model should be trained on the train subset and then be tested on the test subset, which is not used in the training process. The employed strategy for this stage is five-fold cross-validation, where the whole dataset is split into five equal folds. Then in five different experiments, the network is trained using only three folds, validated in one of another folds, and the last one is kept to test the network, whereupon in all the five steps, the test dataset is unique and different. This step confirms the independency of the method on the selected test subsample.

For this reason, we used a Core i7 processor system empowered by a CPU of 3.4 GHz and RAM of 32 GB besides Nvidia GeForce GTX 1080 GPU by the library of Tensorflow.

Mean, SD and frequency are reported for all the quantitative continuous and qualitative nominal variable parameters.

\section{Results}\label{res}
\begin{figure}[!ht]
	\centering{\includegraphics[scale=0.5]{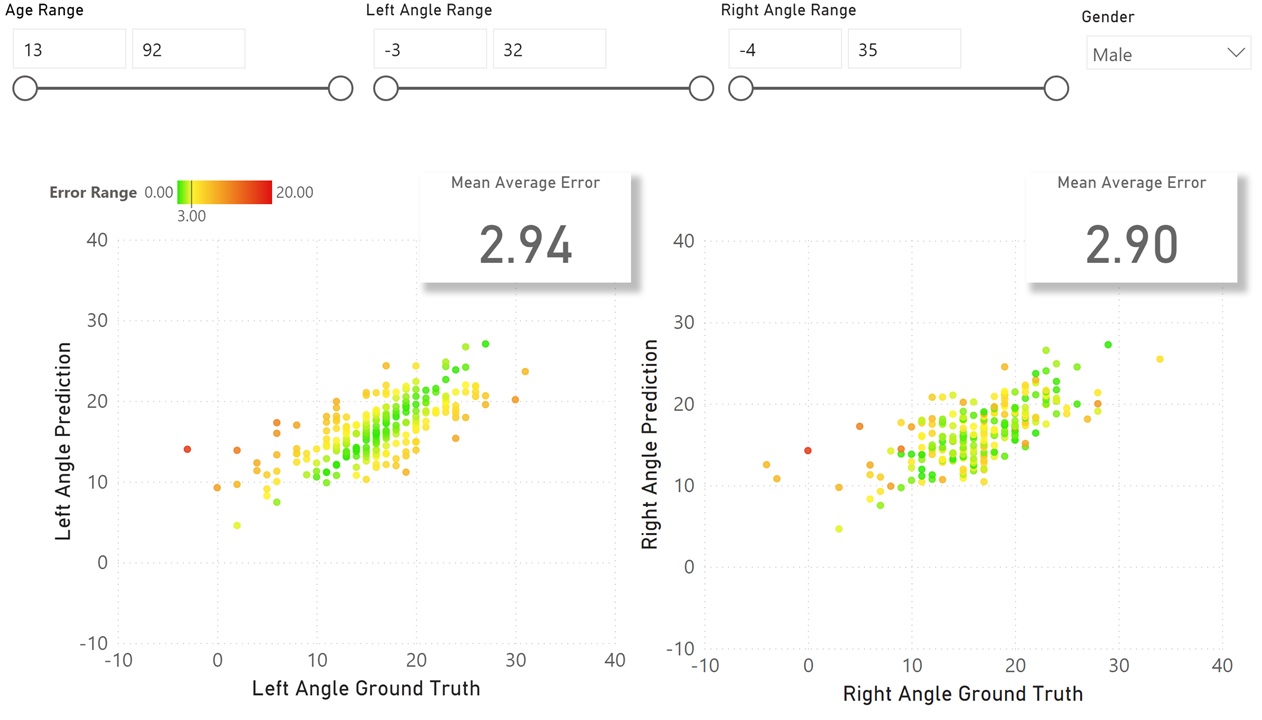}}
	\caption{The scatter plots of both angles of ground truth (CT scan) and predicted by proposed methods for men aged 13-92 years old.\label{pic:male_13-92_2}}
\end{figure}

\begin{figure}[!ht]
	\centering{\includegraphics[scale=0.5]{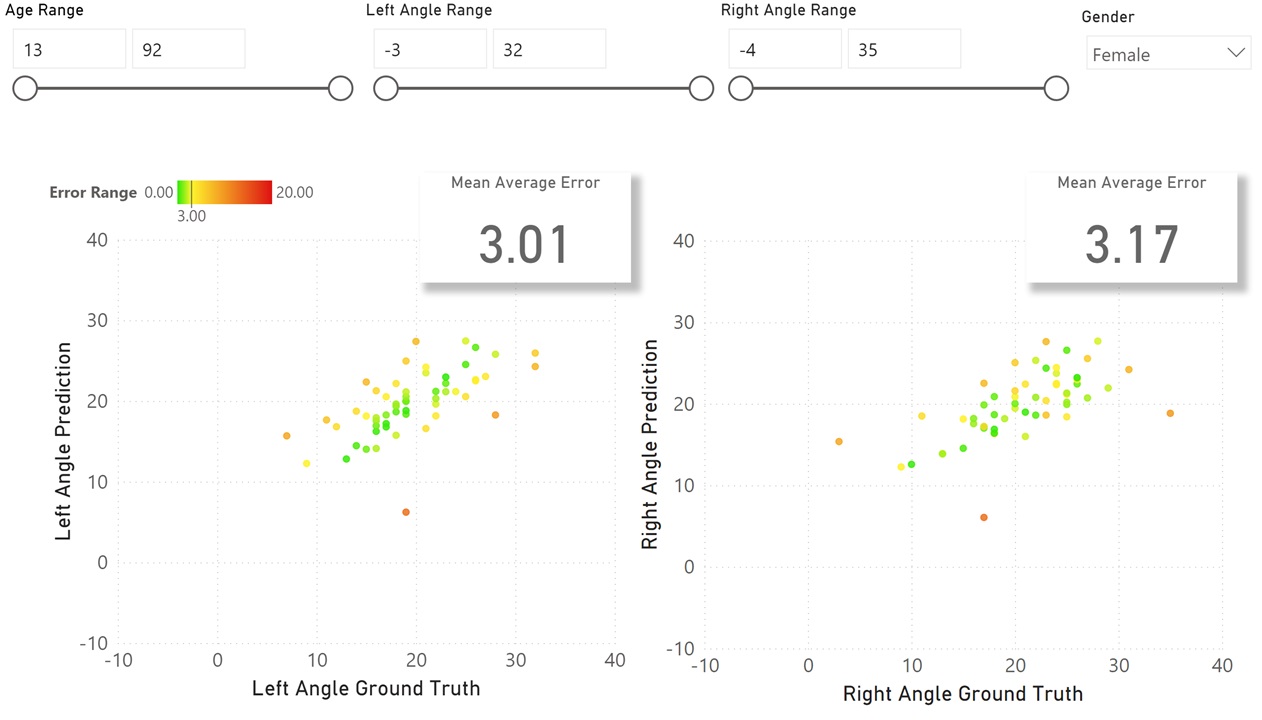}}
	\caption{The scatter plots of both angles of ground truth (CT scan) and predicted by proposed methods for women aged 13-92 years old.\label{pic:female_13-92_2}}
\end{figure}

In the studied population, the values of both hips’ angles computed from CT images increased as people aging, with the mean values of 16.54 and 16.11 (right and left angles) for men and 20.61 and 19.55 degrees for women. As our output was the angle of right and left hips, scatter plots of both angles of ground truth (CT scan) and predicted by proposed methods for men and women are depicted in Fig.\ref{pic:male_13-92_2} and Fig.\ref{pic:female_13-92_2}, respectively. The average error in each plot is also reported with less than 3 degrees as accurate, between 3 to 6 degrees as moderate, and more than 6 degrees as poor.

\begin{figure}[!ht]
	\centering{\includegraphics[scale=0.5]{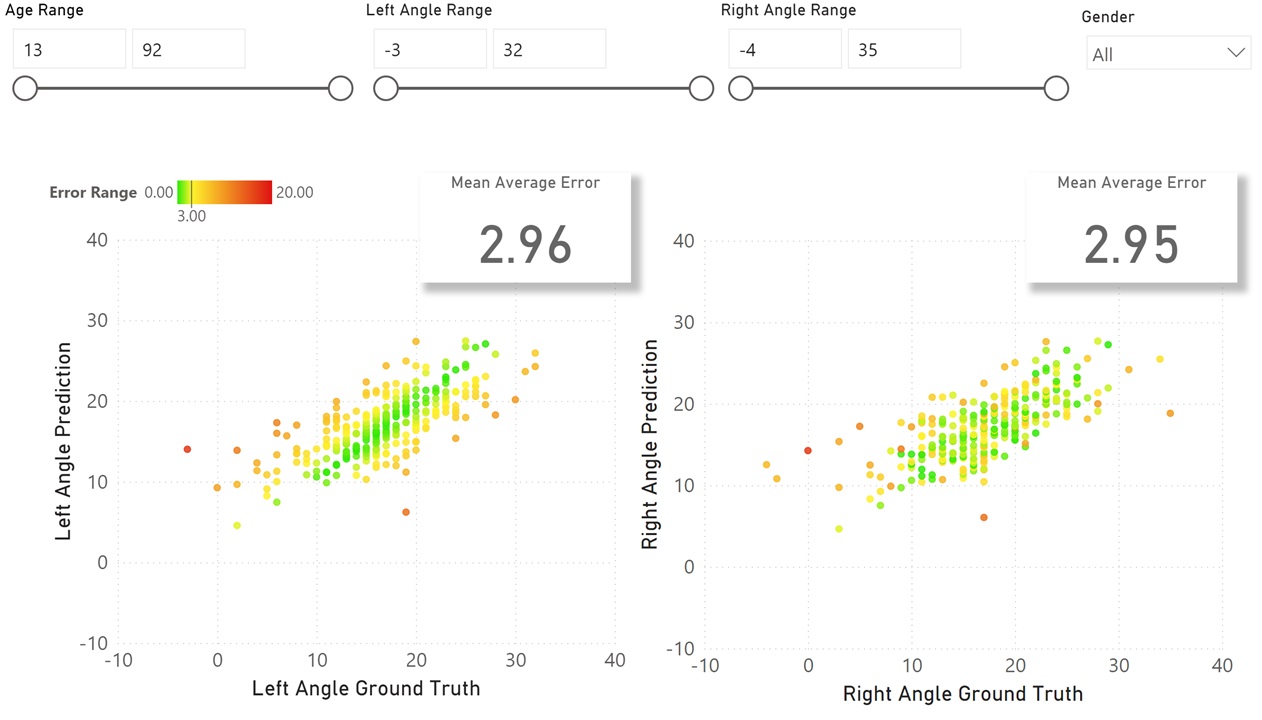}}
	\caption{The scatter plots of both angles of ground truth (CT scan) and predicted by proposed methods for all the population aged 13-92 years old.\label{pic:all_13-92_2}}
\end{figure}

In both measurements of left and right angles for the male population, the average error proves our accurate prediction (green range of error). As shown in Fig.\ref{pic:GenAge} and table \ref{tab:gender} the values of angles in our dataset are limited for women and so the results are situated in the moderate range. Also Fig.\ref{pic:all_13-92_2} shows the scatter plot of all the population with an accurate range of error in both hips.

These results are also reported in table \ref{tab:folds}. It can be seen that the low error of prediction in the accurate region is evidence of the power of the proposed method in angle prediction of both hips in average.

\begin{table}[!t]
	\caption{The computed error in both hips angle prediction for 5 different folds and on average, for male, female and both gender.}\label{tab:folds}
	{\begin{tabular}{|l|c|c|c|c|c|c|c|c|} \cline{1-9}
	&&&\multicolumn{2}{c}{Male}&\multicolumn{2}{|c|}{Female}&\multicolumn{2}{c|}{Both Gender} \\
	\cline{4-9}
	&\#Men&\#Women& Left-Angle & Right-Angle & Left-Angle & Right-Angle & Left-Angle & Right-Angle \\
	&&& -Error &-Error & -Error & -Error & -Error & -Error \\
	\hline
	Fold1 & 51 &  9 & 3.16$\pm$2.60	&3.10$\pm$2.50&3.21$\pm$1.98 &2.59$\pm$2.71&	3.17$\pm$2.50 &3.02$\pm$2.51\\
	Fold2 & 48 &  12 & 3.14$\pm$2.33	&2.81$\pm$2.71&3.59$\pm$2.96 &4.15$\pm$4.41&	3.23$\pm$2.44 &3.08$\pm$3.12\\
	Fold3 & 50 &  10 & 2.93$\pm$2.22	&3.23$\pm$2.55&1.74$\pm$1.21 &3.10$\pm$1.85&	2.73$\pm$2.12 &3.21$\pm$2.43\\
	Fold4 & 48&	12&	2.78$\pm$3.07&	2.70$\pm$2.54&	3.24$\pm$3.04&	2.78$\pm$3.30&	2.87$\pm$3.04&	2.72$\pm$2.68\\
	Fold5&	47&	13&	2.69$\pm$2.31&	2.63$\pm$2.12&	3.11$\pm$3.47&	3.08$\pm$2.99&	2.78$\pm$2.58&	2.73$\pm$2.32\\
	\hline
	Average & & & 2.94$\pm$2.51 &2.89$\pm$2.48 &3.01$\pm$2.72 &3.14$\pm$3.05 &2.96$\pm$2.55 &2.31$\pm$2.61\\
	\bottomrule
	\end{tabular}}{}
\end{table} 

\begin{figure}[!ht]
	\centering{\includegraphics[scale=1]{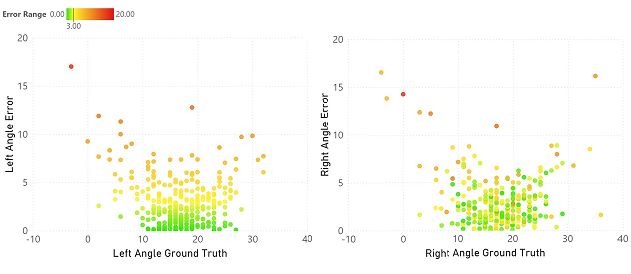}}
	\caption{The average error of right and left angles for the whole dataset.\label{pic:AveError_2}}
\end{figure}

The average error of right and left angles for the whole dataset is exhibited in Fig.\ref{pic:AveError_2}. These displayed charts are also the proof of the ability of the proposed method in recognition of the version in which the majority of errors are less than 6 degrees. 

\begin{figure}[!ht]
	\centering{\includegraphics[scale=0.8]{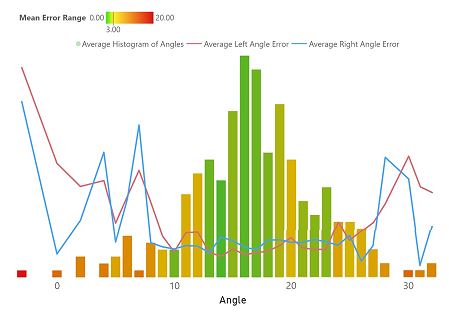}}
	\caption{The error of predicted angles for both hips and the average histogram of angles.\label{pic:hipserror_2}}
\end{figure}

The error of predicted angles for both hips and the average histogram of angles in Fig.\ref{pic:hipserror_2} illustrates the ratio of frequency and average error of angles in which the average error gets lower as the frequency of angles gets higher. The color of bars in the histogram shows the average error of both hips according to the mean average range.

\begin{figure}[!ht]
	\centering{\includegraphics[scale=0.75]{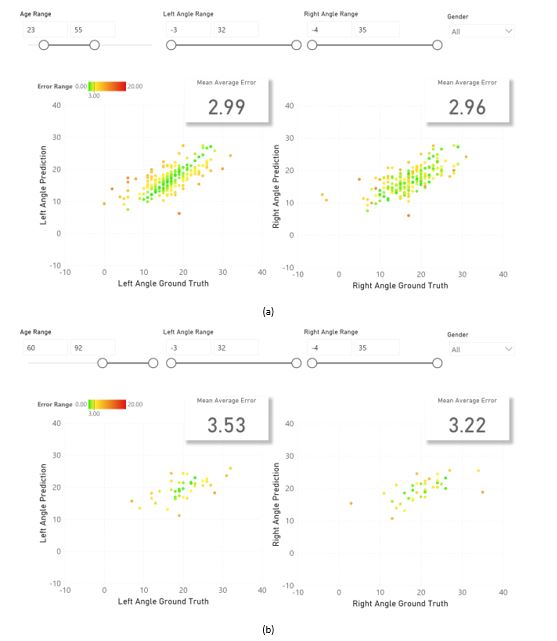}}
	\caption{The scatter plots of both angles of ground truth (CT scan) and predicted by proposed methods for the whole population aged (a) 23-55 and (b) 60-92  years old.\label{pic:all-2goups}}
\end{figure}

As you can see in Fig.\ref{pic:GenAge}, the majority of people are aged 13 to 57 years old. Fig.\ref{pic:all-2goups} reports the scatter plots of both angles of ground truth (CT scan) and predicted by proposed methods for the whole population aged 23-55 and 60-92 years old. As we expected, the more number of people, the lower average error is reported for both hips angles prediction.

All of our results (Figures \ref{pic:male_13-92_2}-\ref{pic:all-2goups} and table \ref{tab:folds}) prove the ability of the proposed method for predicting right and left angles of hips, leads us to achieve the acceptable error. 
\clearpage
\section{Discussion}\label{dis}
The orientation of the acetabulum (version) is a significant feature in the biomechanics \cite{ref:6} and pathologies of the hip joint. Determination of acetabulum version is an essential factor in both predicting and following up the outcome in THA in which malposition of the cup is associated with impingement, dislocation, severe wear, and reduction in the range of motion, force patients for revision surgeries. 

Assessment of the orientation of acetabulum has been done by 3-D imaging systems, CT-scan/MRI, and AP pelvic X-rays. The CT scan of the whole body is presented as the gold standard method. The cheap and safe AP radiograph systems are the initial choice of imaging tool in patients with hip pain.
In this study, we used AP radiographs as an excellent method for assessment of the orientation of the acetabulum. Based on applying DL, the subset of AI, and its development in image diagnosis problems including radiology, the optimum performance of CNN-based algorithm is reported in a wide range of literature. We also examined the performance of attention on pre-trained VGG16 for the bone age model for our purpose of anatomical acetabular version measurement using available AP images, reducing the need for CT scans.

The input of the model was AP pelvic radiographs of patients and as the output the angle of the right and left hips were computed. Our reported results demonstrate the ability of the model for version estimation. Thanks to AI capability, acetabular version measurement can be done by only using anterior-posterior pelvic X-rays, eliminating the need for high radiation and expensive CT scans.

As a limitation in our study, despite the bold advantages (no need for CT images, so anatomical version estimation can be done based on only cheap, low-radiation, and accessible AP radiographs), the AP pelvic radiographs are not really standard. To overcome the resulting error, we considered the minimum properties of a standard photo, namely (i) symmetric obturator foramina, (ii) symmetric iliac wings, and (iii) sacrum and symphysis pubic being at the same line.

\section{Conclusion}\label{con}
To conclude, in our current study, we proposed a method to estimate version in hip disorders and THA. Our results show that suggested algorithm, applying pre-trained vgg16 on the AP images of the pelvis of patients followed by the attention model, can assess anatomical version accurately. We achieved it by using only AP radiographs in which obviate the necessity for CT. The applied technique we used for estimating anatomical acetabular version (based on AP pelvic images using DL approaches), to the best of authors' knowledge, has not been published yet.

\section{Conflict of interest statement}
The authors declare that there are no conflicts of interest.

\printcredits

\end{document}